\documentclass[
prb,aps
 ,amsmath,amssymb
 ,twocolumn,a4paper,twoside,
 floatfix
]{revtex4-2}
\usepackage[english]{babel}
\usepackage{graphicx}
\usepackage{placeins} %
\usepackage{dcolumn}
\usepackage{bm}
\usepackage{float}
\usepackage{verbatim}
\usepackage[hidelinks]{hyperref}

\usepackage{mathabx} 

\usepackage[mathlines,pagewise,modulo]{lineno}

\usepackage{ulem}
\usepackage{xcolor}
\definecolor{eloicyan}{rgb}{0.,0.64,0.84}
\definecolor{darkgreen}{rgb}{0.,0.5,0.}
\definecolor{lightgreen}{rgb}{0.,0.76,0.76}
\definecolor{darkblue}{rgb}{0.1,0.2,0.46}
\definecolor{lightred}{rgb}{1.0,0.3,0.3}
\definecolor{darkred}{rgb}{.7,0.1,0.1}
\definecolor{c1}{rgb}{1.0,0.5,0.3}
\definecolor{c2}{rgb}{.5,0.1,0.2}
\definecolor{c3}{rgb}{.5,0.5,0.5}
\definecolor{lilac}{rgb}{0.6,0.2,0.7}

\newcommand\redout{\bgroup\markoverwith{\textcolor{lightgreen}{\rule[0.5ex]{2pt}{.7pt}}}\ULon}

\newcommand\cyanout{\bgroup\markoverwith{\textcolor{eloicyan}{\rule[0.5ex]{2pt}{.7pt}}}\ULon}

\newcommand\Lilacout{\bgroup\markoverwith{\textcolor{lilac}{\rule[0.5ex]{2pt}{.7pt}}}\ULon}

\newcommand*{\eff}{{\rm eff}}
\newcommand*\TAC{\ensuremath{T_{\rm AC}}}
\newcommand*\TMC{\ensuremath{T_{\rm MC}}}

\renewcommand{\vec}[1]{\boldsymbol{\mathbf{#1}}} 
\newcommand{\unitvec}[1]{\hat{\boldsymbol{\mathbf{#1}}}} 

\DeclareMathAlphabet\mathbfcal{OMS}{cmsy}{b}{n} 

\begin{document}


\title{
Size-dependent mobility of skyrmions beyond pinning in ferrimagnetic GdCo thin films
}

\author{L\'eo Berges}
\author{Eloi Haltz}
\author{Sujit Panigrahy}
\author{Sougata Mallick}
\author{Rapha{\"e}l Weil}
\author{Stanislas Rohart}
\author{Alexandra Mougin}
\author{Jo\~ao Sampaio}
 \email{joao.sampaio@universite-paris-saclay.fr}
\affiliation{
Universit\'e Paris-Saclay, CNRS, Laboratoire de Physique des Solides, 91405 Orsay, France
}

\date{\today}

\begin{abstract}
    Magnetic skyrmions are swirling magnetic textures that can be efficiently driven with spin-orbit torques with a deflected trajectory. However, pinning slows skyrmions down and alters their trajectory, which prevents a quantitative comparison to analytical models.
    Here, we study skyrmions driven by spin-orbit torques at room temperature in ferrimagnetic GdCo thin films, an amorphous material with low pinning.
    Above a sharp current depinning threshold, we observe a clearly linear velocity increase with current that extrapolates to zero and a constant deflection angle, reaching high velocities up to 200~m/s.
    The mobility increases and the depinning threshold current decreases with the skyrmion diameter, which we vary using an external magnetic field.
    An analytical model based on the Thiele equation quantitatively reproduces these findings with a single fitting parameter.
    This validates the linear flow regime description and shows, in particular, the important role of skyrmion size in its dynamics.
\end{abstract}

\maketitle

\section{Introduction}
Magnetic skyrmions are particle-like magnetic textures similar to magnetic bubbles~\cite{Kooy1960,Malo1979}, from which they differ by their well-defined chirality and non-trivial topology.
In the last decade, advances in the optimization of interfacial Dzyaloshinskii Moriya interaction (DMI) and of current-induced spin orbit torques (SOTs) have allowed the stabilization of extremely small skyrmions, down to tens of nm, and their efficient driving in thin-film tracks \cite{Woo2016,Jiang2017a,Juge2019b}.
Their chirality is a crucial element to SOT driving, while their non-trivial topology manifests itself by a deflected trajectory.

Three regimes for skyrmion motion can be distinguished with increasing driving current: a pinning regime, where the effects of the film inhomogeneities are strong, a linear flow regime, where the skyrmion propagates while conserving its shape, and a non-linear flow regime, where the skyrmion shape is deformed.
In the flow regime, analytical and numerical models \cite{Reichhardt2016a,Chen2022} describe the skyrmion dynamical laws and identify the main governing parameters: the current density ($J$), the skyrmion size, and the material's angular momentum density ($L_S$).
The skyrmion velocity is expected to increase linearly with current density while the deflection should stay constant \cite{Tomasello2014c,Buttner2015}. As a function of skyrmion size, the velocity is expected to increase whereas the deflection should decrease.  The deflection should be proportional to $L_S$, which has motivated an interest for antiferromagnetic or ferrimagnetic materials where $L_S$ can be zero and the skyrmion deflection suppressed.

However, in most studied systems, the effects of pinning are significant even at highest applied current densities. Pinning slows down skyrmions, stops their motion below a certain threshold, and interferes with their deflection and shape \cite{Reichhardt2016a,Legrand2017a}.
The flow regime, where the skyrmion should propagate with negligible effects from pinning and thus with high mobility, is often inaccessible due to the limitations that Joule heating imposes on the applied current density.
Experimental observations, therefore, agree only partially with these predictions or are incomplete. For example, instead of being constant, the deflection has most often been observed to increase with current eventually saturating at the highest currents \cite{Yu2016,Jiang2017a,Juge2019b,Dohi2019}.
The velocity was observed to increase linearly with current in some experiments \cite{Woo2018b,Litzius2017}, but often follows a shifted linear law \cite{Juge2019b,Hrabec2017c,Woo2016,Litzius2020} or even an highly non-linear variation with current, quantitatively slower than the velocity expected for the linear regime \cite{Jiang2015,Jiang2017a,Dohi2019,Zeissler2020}.
The dependence of the deflection angle on the diameter was observed in some cases \cite{Jiang2017a,Litzius2017} but not in others \cite{Zeissler2020}.
The dependence of the velocity with diameter was found to be constant in ref. \cite{Tan2021}.
The variation of the deflection angle with $L_S$ was observed for domains with chiral domain walls~\cite{Hirata2019}, but not for isolated skyrmions.

Here, we study the SOT-driven dynamics of skyrmions in a thin film of GdCo, a rare earth/transition metal (RETM) ferrimagnetic alloy.
The high sensitivity of magnetization ($M_S$) to temperature and alloy composition in RETMs \cite{Hansen1989} is used to tune the stability of skyrmions near room temperature.
The very low pinning of this amorphous material allows the study the skyrmion dynamics in the flow regime and to attain higher mobility than what was observed in other thin films.
The variation of the skyrmion mobility and deflection with current density and skyrmion diameter is studied. These observations are compared quantitatively to an analytical model based on the Thiele equation. We also study the pinning threshold and its evolution with the skyrmion size, which is compared to previous predictions based on simulations of inhomogeneous systems.

\section{Results}

\begin{figure}[!htp]
    \centering
    \includegraphics[width= \columnwidth]{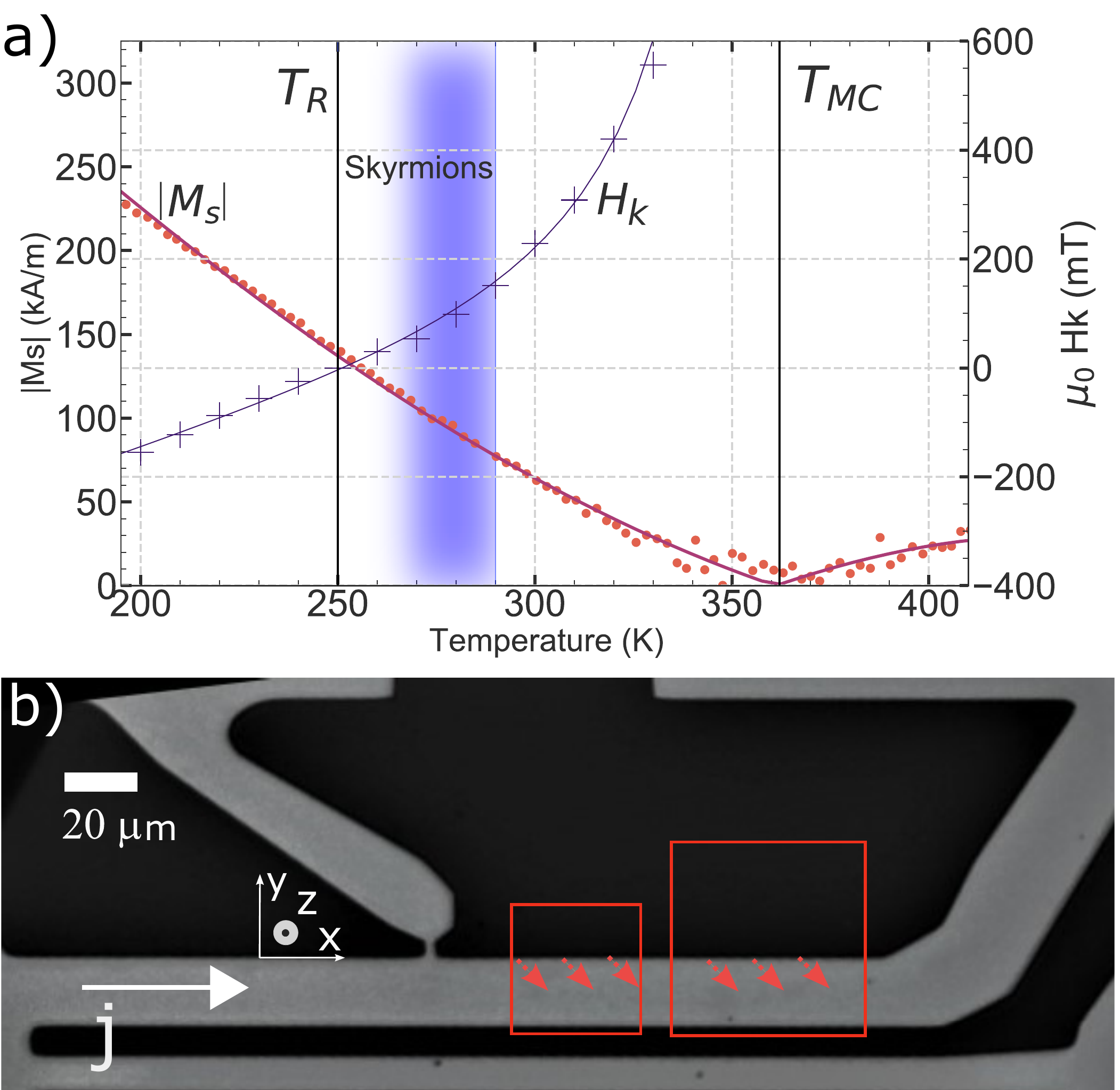}
    \caption{
    a) Saturation magnetization $|M_{S}|$ (red dots measured by SQUID, line is given by the mean field model) and effective anisotropy field $H_{k}$ (measured from AHE loops). The temperature range highlighted in blue shows where skyrmions were observed at low or zero magnetic field.
    b) Optical image of the device. The magnetic track appears in light gray. The two red squares delimit the field of view of the skyrmion tracking measurements. The red arrows indicate the typically observed nucleation side and propagation direction of skyrmions. }
    \label{fig:charact}
\end{figure}

A Ta(1)/Pt(5)/Gd$_{0.32}$Co$_{0.68}$(5)/Ta(3) (thicknesses in nm) thin film was deposited by e-beam evaporation in a ultra-high-vacuum chamber \cite{Haltz2018a} on a Si/SiOx (100~nm) substrate, and patterned as 20–$\mu$m-wide tracks (Fig.~\ref{fig:charact}b).
The magnetization of the film ($M_S$) versus temperature ($T$) was measured by SQUID (Fig.~\ref{fig:charact}a), showing a strong variation typical of RETM ferrimagnets~\cite{Hansen1989}. It shows a magnetic compensation temperature $\TMC{}=~360$~K, at which the magnetic moments of the two antiferromagnetically-coupled sublattices (Gd and Co) balance out and $M_S=0$. Analogously, the angular momenta of the sublattices  balances out at the angular momentum compensation temperature, \TAC{} which we expect to be about 425~K.
The effective anisotropy field ($H_{k}$), shown in Fig.~\ref{fig:charact}a, was extracted from fits of anomalous Hall effect (AHE) hysteresis cycles taken with a tilted field~\cite{SI}.
The fitted $H_{k}$ increases with temperature up to 360~K, and is negative below 250~K. The point of $H_K=0$ defines the reorientation temperature $T_R$, where the anisotropy changes from a hard ($T<T_R$) to an easy ($T>T_R$) perpendicular axis.

\begin{figure}[!htp]
    \centering
    \includegraphics[width =\columnwidth]{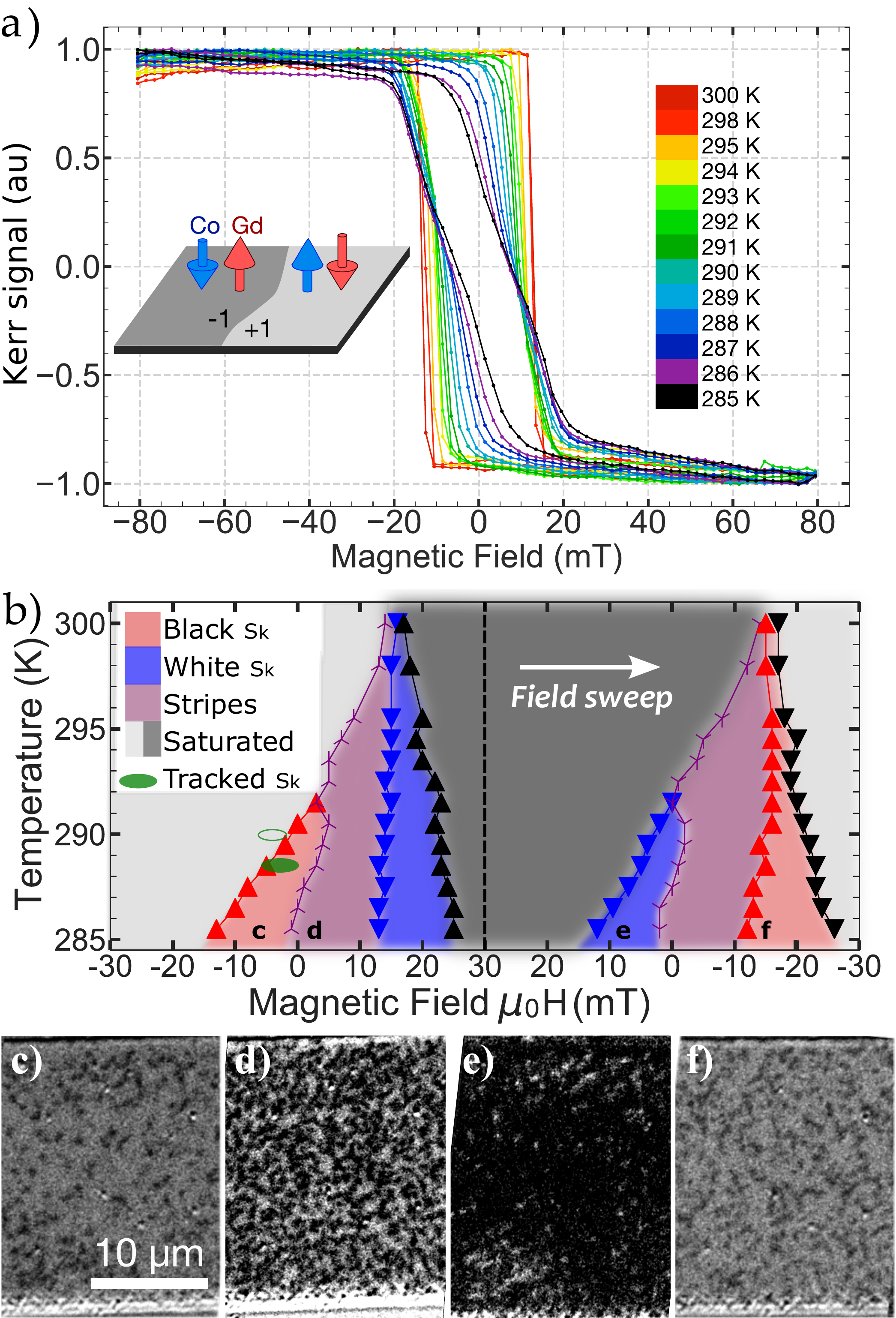}
    \caption{
    a) MOKE hysteresis loops near the temperature range of skyrmion stability.
     Inset: Diagram of the orientation of the Co and Gd sublattices in the domains and corresponding Kerr signal.
    b) Phase diagram of magnetic textures during field hysteresis loops from -30 to 30 mT  and back to -30 mT. The phases and their transition fields are defined in the text.
    Red, blue and black triangles and the purple stars correspond to the phase transition fields in the 13 measured field sweeps from which the phase borders (lines) are deduced.
    The green empty and filled regions correspond to the two temperatures of the skyrmion dynamics measurements.
    c-f) MOKE differential images at $T=285.5$~K and at the magnetic fields indicated in b).   }
    \label{fig:diag}
\end{figure}

The hysteresis cycles with perpendicular field taken at different temperatures below \TMC{} were measured by magneto-optical Kerr effect (MOKE) and are shown in Fig.~\ref{fig:diag}a.
Because the measured MOKE signal is mostly due to the Co sublattice~\cite{Hansen1989,Haltz2018a} and $T<\TMC{}$, a positive saturation is observed at negative applied magnetic field  and corresponds to a Co magnetization pointing along $+\unitvec{z}$ (in the coordinate system of Fig.~\ref{fig:charact}b) and a net magnetization along $-\unitvec{z}$. By convention, we describe the orientation of the magnetic film using the orientation of the Co \cite{Wangsness1953}.
Slightly above $T_{R}$ and below 290~K, the hysteresis cycles present a peculiar shape with low remanence which is a sign of a multi-domain state containing textures such as stripes or skyrmions.

The MOKE images taken during the hysteresis loops reveal different types of textures, which we classify in four phases, shown in Fig.~\ref{fig:diag}b: saturated, black or white skyrmions  (with the cobalt moment at the center along $-\unitvec{z}$ or $+\unitvec{z}$, respectively), and stripes.
The transition from skyrmions to stripes (shown as lines and symbols in Fig.~\ref{fig:diag}b) is smooth. To distinguish them, we extract the shapes of the domains from the images using a watershed segmentation algorithm \cite{Kornilov2018} and calculate their surface-to-perimeter ratios. A threshold was chosen to classify the textures either as stripes (with higher ratio) or skyrmions (with lower ratio). Similar results were obtained by using the eccentricity of the best-fitting ellipse (with a different threshold).

Starting with a saturated state at $-30$~mT and increasing the field, we observe  skyrmions (black dots in the MOKE image of Fig.~\ref{fig:diag}c) above a temperature-dependent nucleation field (first red triangles). Then a stripe, labyrinth-like phase develops, and becomes denser with increasing positive field (Fig.~\ref{fig:diag}d). The white stripes are then compressed into sparse white skyrmions (Co moment along $+\unitvec{z}$), before reaching the full saturation of the sample. A symmetric evolution is observed along the opposite field sweep (Fig.~\ref{fig:diag}e,f).

Skyrmions are only observed above $T_R$ and on a narrow range of  field, which is narrower for higher temperatures.
When dipolar energy is important and the skyrmion diameter is large, its stability can be understood by using the analysis developed for magnetic bubbles \cite{Kooy1960} and later refined for the skyrmion case \cite{Rohart2013b, Boulle2016, Bernand-Mantel2018b}.
As a first approximation, skyrmions or stripes are expected when the characteristic dipolar length
$l_C=\sigma/(\mu_{0} M_{s}^2)$ (where $\sigma$ is the domain wall energy)
is greater or comparable to the film thickness. This is indeed the case near 290~K (see Appendix~\ref{app:Params}), where spontaneous textures are observed.

\begin{figure}[htp]
    \centering
    \includegraphics[width =0.9 \columnwidth]{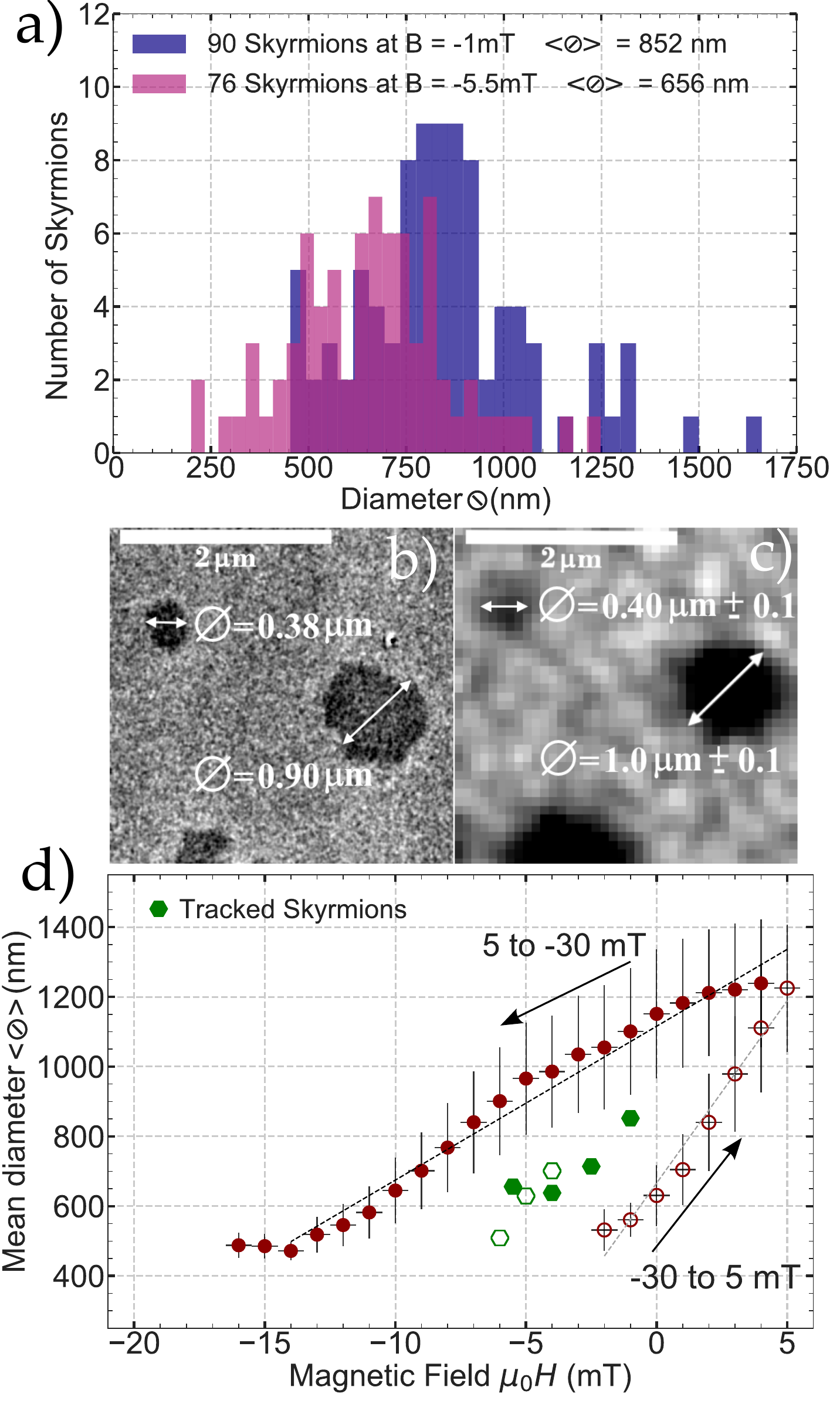}
    \caption{ a) Skyrmion diameter histogram for two magnetic fields at $T=288.5$~K. b) MFM and c) MOKE images of the same area showing two skyrmions. d) Full and empty red circles show the average skyrmion size versus magnetic field applied along +$\unitvec{z}$ for $T = 288.5$~K. Green filled (empty) hexagons show the sizes for the tracked skyrmions at 288.5~K (290~K). The error bars are the standard deviation of sizes measured in the images. Due to the Faraday effect during the field sweep, the binarization threshold is field dependent.}
    \label{fig:size}
\end{figure}

The skyrmions show different sizes and shapes. To analyze them, we define the skyrmion diameter, $\diameter$, as the one of a circle with the same surface.
The measured $\diameter$ are often close to the optical resolution. To validate our procedure, MOKE images were compared with high-resolution measurements of magnetic force microscopy (MFM) of the same skyrmions, shown in Figs.~\ref{fig:size}b,c.
Depending on the algorithm threshold parameter, we obtain $\diameter  = 0.4 \pm 0.1$~$\mu$m for the small skyrmion  in the image (compared to 380~nm in MFM) and   $ \diameter  = 1.0 \pm 0.1$~$\mu$m for the larger skyrmion (900~nm in MFM), which validates the use of MOKE at least down to 0.4~$\mu$m.

Fig.~\ref{fig:size}a shows the histograms of skyrmion diameters at two different applied fields (at $T=288.5$~K). At $B=-1$~mT, an average diameter of $\langle \diameter \rangle =852$~nm and a standard deviation of 230~nm are obtained.
The rather large size distribution is explained by the effects of pinning \cite{Gross2018} coupled to the very low domain-wall energy near $T_R$, which lead to a very shallow energy versus size profile \cite{Bernand-Mantel2018b}. The size also shows a hysteretic behavior with field, shown in Fig.~\ref{fig:size}d for black skyrmions during a minor hysteresis loop. This is also compatible with the effects of pinning \cite{Gross2018,Juge2018,Zeissler2017}: during the positive field sweep, the skyrmion enlarges but does not attain the equilibrium diameter due to pinning. The opposite happens when the field decreases, which results in the shown hysteretic behavior, where the equilibrium diameter is inside the hysteretic gap. A clear decrease of the average size with field is obtained. A numerical micromagnetic study reproduces this trend in field (see Fig.~\ref{fig:radius} in Appendix \ref{app:Simus}).

\section{Dynamics under current}
We studied the skyrmion current-driven dynamics in a 20-$\mu$m-wide magnetic track (Fig.~\ref{fig:charact}b) and focused on sparse skyrmions after saturating the sample with a negative field ($T=285.5$~K; see SI~\cite{SI} for 290~K). 10-ns electrical pulses of various current densities $J$ were applied and the size and position of the skyrmions were tracked with MOKE images. After the pulse, skyrmions were nucleated most often on the edges, which act as defects, and most often on the side favored by the Oersted field~\cite{SI}.
Note that skyrmions can be nucleated by current even in states that were initially saturated after the field sweep  (green ellipses in Fig.~\ref{fig:diag}b).
The velocity is calculated from the skyrmion center position before and after the pulse.
A single pulse was injected between images for the fastest skyrmions, whereas 5 to 1000 pulses were injected for the slowest skyrmions (at a low repetition rate, 50~Hz, ensuring no cumulative Joule heating effect).

We have tracked a large number of skyrmions using a semi-automated process consisting in manually identifying the same skyrmion in successive images and automatically measuring their size, velocity, and deflection angle.
Some 800 skyrmions were tracked along 2500 images, giving an average of 3 images per determined velocity at two different temperatures for several magnetic fields.
This large number of skyrmions drastically improves the precision of the data analysis even with a significant dispersion of skyrmion size and velocity.

\begin{figure}[!ht]
\centering
\includegraphics[width=\columnwidth]{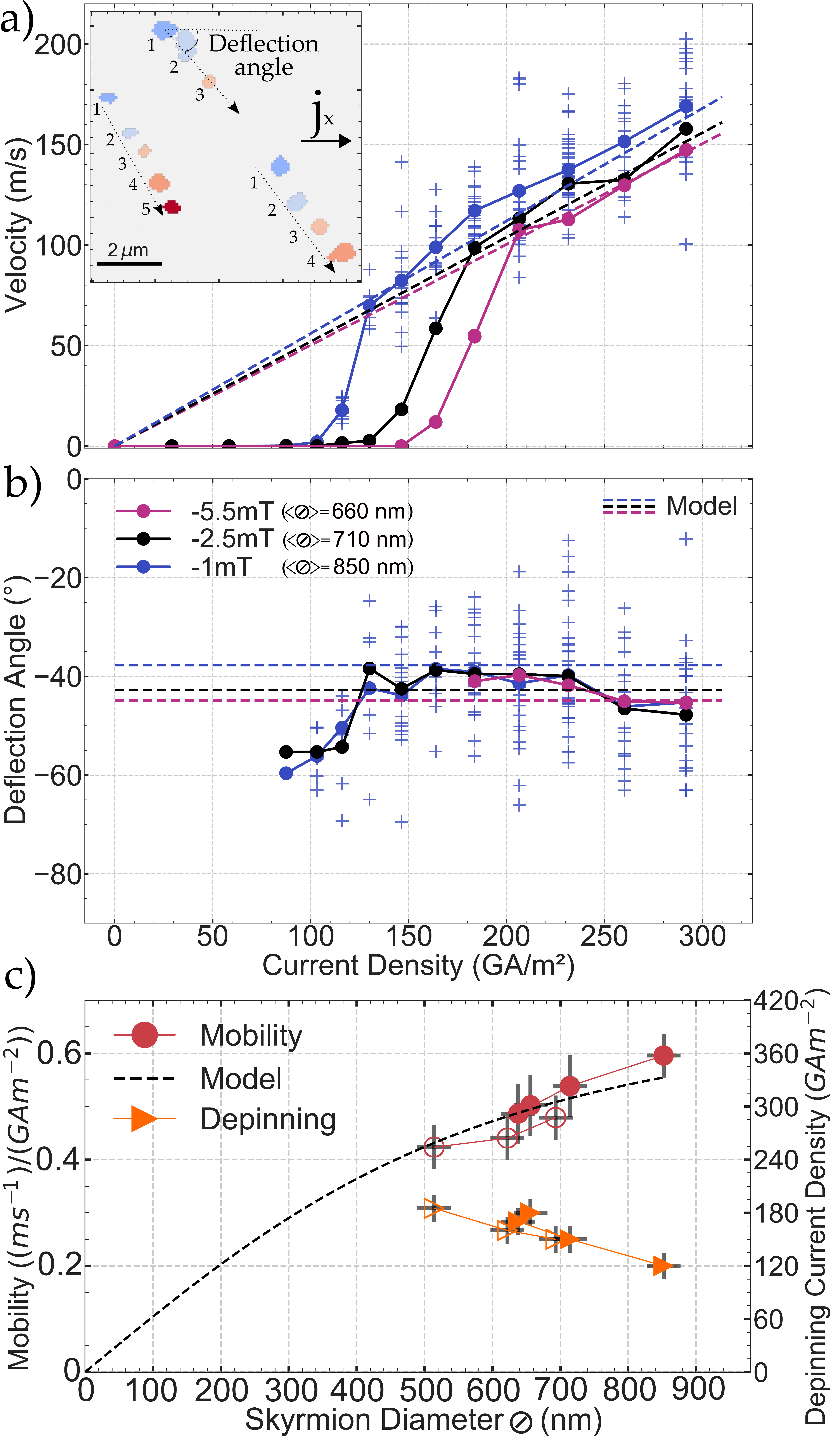}
\caption{ a-b) Skyrmion velocity and deflection angle versus current density, obtained for three different magnetic fields. The line and circles are the mean values of all tracked skyrmions, and the crosses are the measurements for individual skyrmions (only shown for $-1$~mT for clarity). The inset in a) shows the superposition of the binarised images after 5 successive 10-ns-long pulses of 210~GA/m$^2$ showing three propagating skyrmions. c) Skyrmion depinning current density and mobility versus skyrmion diameter measured at 288.5~K (filled markers) and 290~K (empty markers). The dashed lines in a-c correspond to the values calculated by the model.}
\label{fig:VelocityAndDeflection}
\end{figure}

The inset in Fig.~\ref{fig:VelocityAndDeflection}a shows the parallel trajectories of three black skyrmions (with Co moment along $-\unitvec{z}$) over five successive superimposed images.
Note that some skyrmions disappear and are not present in the last images. Indeed, most skyrmions disappear after a few frames, and only a few cross the entire track especially beyond the region where the Oersted field is opposite to the skyrmion core \cite{SI}.
Skyrmions move along the current direction ($\unitvec{x}$), opposite to the flow of electrons, which is coherent with the SOT and chirality given by a Pt buffer layer. The presence of SOT is confirmed by torque measurements \cite{SI}.
There is a significant deflection towards $-\unitvec{y}$, that we attribute to the gyrotropic force related to the skyrmion topology \cite{Thiele1974}. Note that it is opposite to that of a skyrmion with same cobalt core orientation in a pure cobalt film \cite{Hrabec2017c,Jiang2017a}.
Indeed, the gyrotropic deflection is expected to be proportional to the $L_S$ of material, which is negative in this GdCo film below \TMC{}, opposite to that of a pure Co film.

Figs.~\ref{fig:VelocityAndDeflection}a and b show the skyrmion velocity and deflection angle versus current density, $J$, at three values of applied magnetic field. We observe, at low $J$, a hopping regime of very few moving skyrmions with a badly defined deflection angle and very slow velocity. This is followed by an abrupt transition at a depinning current density $J_D$  above which most skyrmions propagate and obey a linear velocity regime that extrapolates to zero. These skyrmions move up to 170 m/s (on average) for the largest current density of $J~=~290$~GA/m$^2$ and $B=-1$~mT, resulting in a mobility of 0.59~ms$^{-1}$/GAm$^{-2}$.

 Skyrmions display a large deflection angle of $-40^{\circ}$~$\pm 10^{\circ}$ once the linear regime is reached.
 During the depinning regime, the angle is found to be larger. However, only a few skyrmions moved in these conditions and were mostly located near the edge of the track, which may strongly bias the measurement of the angle.
 The angle seems also to increase at the largest current densities, which may be due to an effect of shape distortions \cite{Litzius2020}.

 The mobility and $J_D$ are observed to vary with the applied magnetic field: the lowest $J_D=125$ GA/m$^2$ and highest mobility are observed for the lowest field ($-1$~mT).
 No clear effect of the field on the deflection angle can be discerned.

The average size of the tracked skyrmions (green hexagons in Fig.~\ref{fig:size}d) decreases with increasing magnetic field, as expected. Also, the size of the tracked skyrmions lay within the hysteretic gap of the sizes measured during field sweep, which indicates that they are close to the equilibrium diameter for that applied field (as the equilibrium diameter is inside the hysteretic gap). Note, however, that the images capture the skyrmion tens of ms after the pulse, and so the measured size may differ from the size during motion.
We observe the lowest depinning current $J_D$ and the highest velocity for the smallest applied magnetic field, i.e., for the largest skyrmions. The relation between the dynamic features and the magnetic field might thus be attributed to the change in skyrmion size (Fig.~\ref{fig:VelocityAndDeflection}c).
As for depinning, previous simulation studies of skyrmions pinned in granular films~\cite{Legrand2017a,Salimath2019} predict that the pinning threshold varies strongly with skyrmion size and attains a maximum when the diameter is comparable to the length scale of the variation of magnetic parameters.
The presented data shows a strong variation of pinning with size in an amorphous film, suggesting that the conclusions of those studies may also be applied to inhomogeneous films other than multi-crystalline materials. The observed variation of $J_D$ with diameter is compatible with a characteristic pinning length smaller than the observed diameters. Although not directly comparable, microscopy cross section observations of a similar material (GdFeCo) deposited in the same conditions \cite{Krishnia2021} showed lateral inhomogeneities with a length scale around 3 nm, suggesting that the pinning length scale of this film is indeed smaller than the observed diameters.

The observed velocity and mobility are significantly higher than what has been observed in other thin films (see reference \cite{Woo2018b} for GdFeCo and, for other materials, e.g., \cite{Juge2019b,Litzius2017,jiang2016,Woo2016}). As we are far from \TAC, this high mobility is not due to the particularities of ferrimagnetic dynamics, but mainly due to the fact that the skyrmions attain the linear velocity regime due to the weak pinning of this material, whereas most other skyrmion propagation observations were performed below or near the pinning threshold. Another factor for the lower mobility reported in previous studies is the smaller skyrmions that were studied, as is discussed in the following section.

\section{Modeling skyrmion dynamics}
 To examine the effect of the size on the skyrmion velocity and deflection, we use a model based on the Thiele equation, which describes the motion of rigid magnetic textures \cite{Thiele1973,Thiele1974}. This model, described in detail in the Appendix \ref{app:Thiele}, yields the skyrmion velocity
\begin{equation} \label{eq:modelVelocity}
    |v| = \frac{v_{0}}{\sqrt{1+\rho^2}} ,
\end{equation}
where $\rho$ is the deflection rate (the deflection angle is $\arctan(\rho)$) and $v_{0}$ is the velocity without deflection, given by
\begin{equation}
    v_{0}  =  \frac{\hbar J \theta_{\rm SHE}}{2 e  L_\alpha t} \frac{f}{d} \hspace{0.5cm} {\rm and} \hspace{0.5cm}
    \rho \equiv \frac{v_{y}}{v_{x}} =  \frac{L_S}{L_\alpha} \frac{n}{d}  ,
\end{equation}
 where $L_S=M_S/\gamma$ is the net angular moment density (negative as $T<\TAC{}$, as discussed earlier),
 $L_\alpha$ is the energy dissipation rate \cite{Haltz2021,VellaColeiro27},
 $\theta_{\rm SHE}$ is the spin Hall angle, $t$ is the film thickness, $e$ is the elementary charge, and $\hbar$ is the reduced Planck's constant.
The parameters $n$, $f$ and $d$ are given by spatial integrals of the skyrmion magnetization profile (see Appendix~\ref{app:Thiele}). In particular, $f$ and $d$ depend on its radius.

In the limit of small radius compared to domain wall width, the model predicts that $v_0 \ \propto\  f/d \ \propto\ R$ and that $\rho \ \propto \  n/d \rightarrow 1$ (its maximum).
In the large radius limit, $v_0$ is independent of the $R$ and $\rho \ \propto\ 1/R$.
Therefore, the velocity is expected to increase with skyrmion size and the deflection angle to decrease.
Therefore, the velocity $v$ (Eq.~\ref{eq:modelVelocity}) is expected to increase and saturate with skyrmion size, while the deflection angle should be constant for small skyrmions and then decrease.
For a given radius, this model, which does not account for pinning, predicts a linear dependence of the velocity with current and a constant deflection angle.

To apply the model to our observations, we characterized the material parameters using  Brillouin light scattering (BLS) measurements~\cite{SI}.
The mobility $\mu = |v|/J$ (Fig.~\ref{fig:VelocityAndDeflection}c) was fitted with this model with only $\theta_{\rm SHE}$ as a free parameter (which is only a scaling factor), found to be
in the range of the experimental value obtained from torque measurements \cite{SI,Hayashi2014a}.
The model predictions with these parameters are shown by the dashed lines in Fig.~\ref{fig:VelocityAndDeflection}. They reproduce the size dependence of the linear velocity regime of skyrmions taken at 288.5 K and 290 K.
The predicted deflection angles are consistent with the data, but the agreement is less satisfactory, in the sense that the expected variation of deflection angle with diameter is not observed in experimental angles. Moreover, for the skyrmion sizes we observe (R/$\Delta$ = 15), we expect the mobility variation with size to be driven only by the variation of $\rho$. Therefore, the variation of mobility implies a variation of deflection that is not observed. This may be due to the dispersion of the measured deflection angle, which can blur the variation with size. The model also predicts the angle to be more dispersed than the velocity $v$ for a given size distribution \cite{SI}. Another cause may be due to the effects of skyrmion-skyrmion interactions, which the model does not account for \cite{Capic2020,Brearton2020,Tan2021}.\\

\section{Discussion and Conclusion}

We observe skyrmions in a RETM ferrimagnetic thin film, at room temperature and close to $T_{R}$ with zero external magnetic field.
Skyrmions are driven by SOT and follow a clear linear regime after a steep depinning threshold, that decreases with the skyrmion size.
The flow regime, beyond the effects of pinning, was observed, with a linear dependence of the velocity extrapolating to zero.
The mobility, its dependence on the skyrmion size, and the trajectory deflection angle were found to be in quantitative agreement with an analytical model based on the Thiele equation, with a single fitting parameter.
This shows that the rigid skyrmion model using the Thiele equation is sufficient and quantitative, as long as the skyrmion is unhindered by pinning and even for the largest current density that was applied.
In particular, the model predicts a strong reduction of mobility at smaller diameters, which is potentially a problem for the scaling down skyrmion devices.

The observed large mobility and low pinning show the promise of RETM ferrimagnets as tunable systems to explore and optimize the complex skyrmion static and dynamical properties.
Indeed, in RETMs, changing temperature or composition changes substantially the net magnetization and angular momentum, which can be used to control the skyrmion size and stability as well as its dynamics. We observed a negative deflection angle in a film below \TAC{}, i.e., with negative net $L_S$, opposite to the deflection of a pure Co film. This supports the prediction that the gyrotropic deflection is determined by $L_S$, and that deflectionless skyrmions should be achievable at \TAC{}.

\begin{acknowledgments}
The authors thank Andr\'e Thiaville for the fruitful discussions and the study of the sample properties by BLS.
This work was supported by a public grant overseen by the French National Research Agency (ANR) as part of the \textit{“Investissements d’Avenir”} program (Labex NanoSaclay, reference: ANR-10-LABX-0035, project SPICY) and by an Indo-French collaborative project supported by CEFIPRA (IFC/5808-1/2017). Magnetometry and Anomalous Hall effect measurements were performed at the LPS Physical Measurements Platform.
\end{acknowledgments}

\appendix

\section{Thiele equation for skyrmion dynamics \label{app:Thiele}}
Under the hypothesis of a rigid (or stationary) profile $\vec{m}(\vec{r},t)=\vec{m}_0(\vec{r}-\vec{v}t)$ where $\vec{m}_0(\vec{r})$ is the rigid texture and $\vec{v}$ its velocity, the skyrmion dynamics can be described with the Thiele collective coordinate model~\cite{Thiele1973,Thiele1974}, obtained by spatially integrating the Landau-Lifshitz-Gilbert (LLG). For a circular N\'eel skyrmion subjected to a force induced by a current $\vec{J}$ along the $x$ direction, the Thiele equation reads~\cite{Thiele1973,Thiele1974}
\begin{equation}
\label{eq:Thiele classique}
    \vec{G} \times \vec{v} - \alpha D \vec{v} + \vec{F} = 0
\end{equation}
where $\vec{G} = L_S t n \unitvec{z}$,
$\alpha D =  L_\alpha t d$ ($\alpha=L_\alpha/L_S$ is Gilbert's damping constant) and
$\vec{F} = -\frac{\hbar}{2e}  \theta_H \vec{J} f$.
The three terms in eq.~\ref{eq:Thiele classique} respectively arise from the precessional, damping and SOT terms in the LLG equation.
The parameters $n$, $d$ and $f$ characterize the texture geometry $\vec{m}_0(\vec{r})$ and are given by
\begin{subequations}
    \label{eq:Thiele parameters ndf}
    \begin{eqnarray}
 n &=& \iint \left( \frac{\partial \vec{m}_0}{\partial x}\times\frac{\partial \vec{m}_0}{\partial y} \right)\cdot\vec{m}_0d^2r\\
 d &=& \iint \left( \frac{\partial \vec{m}_0}{\partial x} \right)^2d^2r\\
 f &=& \iint \left( m_{0,x}\frac{\partial m_{0,z}}{\partial x}-m_{0,z}\frac{\partial m_{0,x}}{\partial x} \right)d^2r.
    \end{eqnarray}
\end{subequations}
They correspond respectively to the texture topology, to the magnetization rotation length-scale, and to the texture chirality.

The solution of Eq.~\ref{eq:Thiele classique} is
\begin{subequations}
\begin{eqnarray}
    v & = & v_{0}/\sqrt{1+\rho^2} \\
    v_y/v_x & = & \rho ,
\end{eqnarray}
\end{subequations}
with
\begin{subequations}
    \label{eq:mu and rho}
    \begin{eqnarray}
    v_{0} &=& \frac{F}{\alpha D}=-\frac{\hbar j\theta_H}{2e \alpha L_s~t}\frac{f}{d}\label{eq:mu}\\
    \rho &=& \frac{G}{\alpha D}=\frac{1}{\alpha}\frac{n}{d}\label{eq:rho},
    \end{eqnarray}
\end{subequations}
where $v_0$ and $\rho$ correspond, respectively, to the velocity along the current direction (i.e. the velocity when $v_y=0$), and to the deflection of the skyrmion.

In a ferrimagnet where the coupling between sublattices is strong, a perfect antiparallel alignment of the two sublattices can be assumed:  $\vec{m} = \vec{m}^{\rm Co} = - \vec{m}^{\rm Gd}$, where we set by convention the effective normalized magnetization parallel to the Co moment. In this case, we can apply this result by using the effective model introduced by Wangsness and others~\cite{Wangsness1953, Haltz2021}.

A two Thiele equation model, developed in \cite{SI}, shows that this effective approach is valid even in the case of imperfect antiparallel alignment.
In the effective model, the material parameters are related to the parameters of each sublattice: the net magnetization is $M_S= M_S^{Co}-M_S^{Gd}$, the net angular momentum is $L_S = L_S^{\rm Co}-L_S^{\rm Gd}$,  the energy dissipation rate is $L_\alpha = \alpha^{\rm Gd}L_S^{\rm Gd}+\alpha^{\rm Co}L_S^{\rm Co}$, and the effective Hall angle is $\theta_{\rm H} = \theta_{\rm H}^{Gd}+\theta_{\rm H}^{Co}$ \cite{Haltz2021} (the parameters labeled 'Co' or 'Gd' refer to the parameters of each sublattice).  The $M_S$ and $L_S$, and their variation with temperature, can be calculated with a mean field model (Fig.~\ref{fig:charact}a). As is common in RETMs, $\theta_H^{\rm Gd}$ can be neglected.
The effective $\alpha$ diverges at \TAC{} ($\alpha=L_\alpha/L_S$), and so it is convenient to
use the always-finite $L_\alpha$ instead.
The $f$, $d$ and $n$ parameters are unchanged as they are purely functions of the texture's morphology. We can then rewrite eqs. \ref{eq:mu and rho} as
\begin{subequations}
	\label{eq:ThieleSolEffPars}
	\begin{eqnarray}
	v_{0}  &=&  -\frac{\hbar j \theta_H }{2 e L_\alpha t}\frac{f}{d}\\
	\rho &=& \frac{L_S}{L_\alpha}\frac{n}{d}.
	\end{eqnarray}
\end{subequations}
This formalism successfully describes the vanishing gyrotropic deflection expected at the angular momentum compensation \TAC{} (where $L_S=0$) \cite{Zhang2016,Woo2018b,Hirata2019}. The velocity modulus displays a maximum at \TAC{} with $|\vec{v}| \approx |v_{0}|(1-\frac12\rho^2)$.

\section{Effect of skyrmion size on Thiele parameters \label{app:ThieleAndSize}}
\begin{figure}[ht]
\includegraphics[width=\columnwidth]{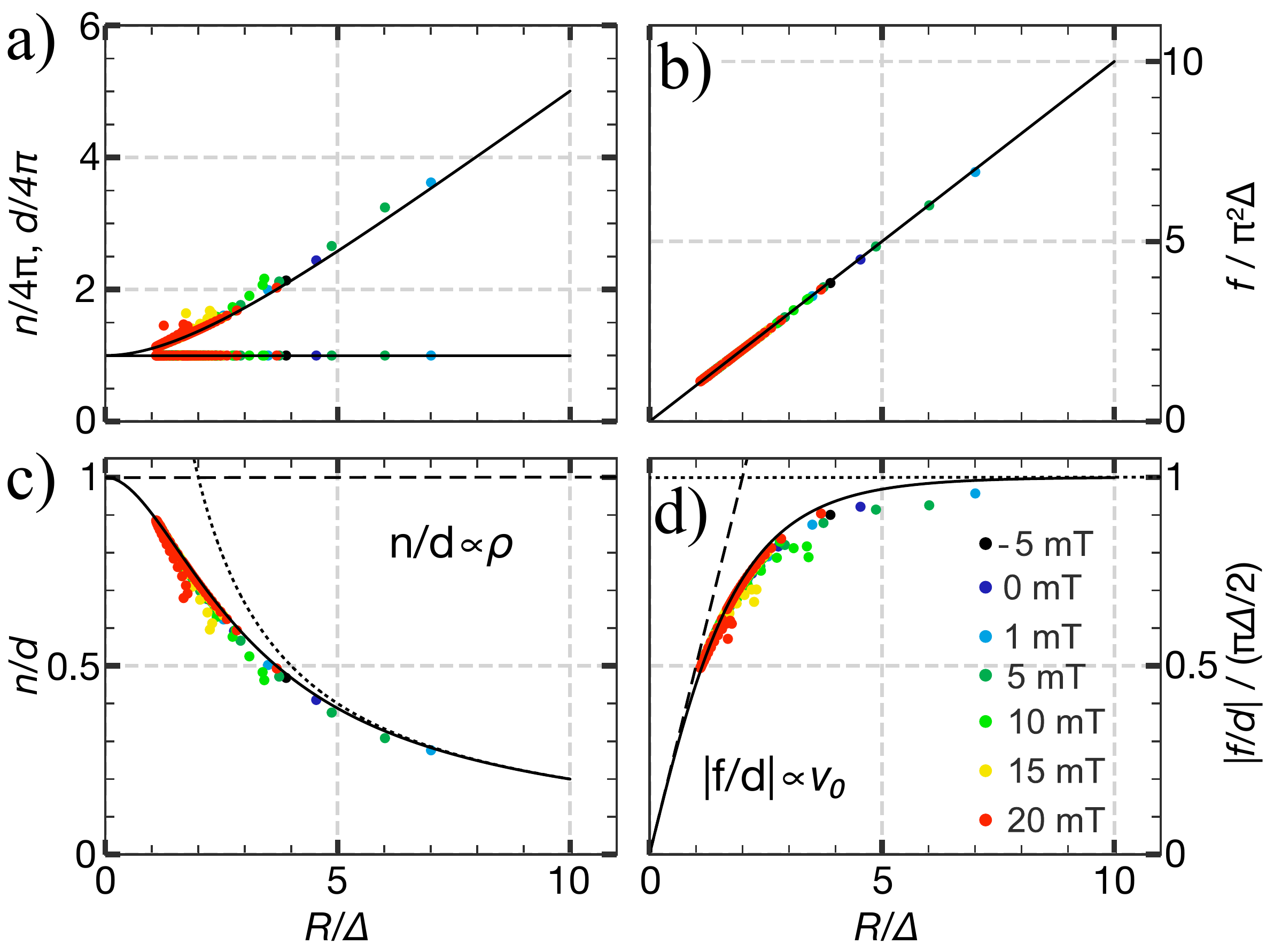}
\caption{
    Geometric components of the Thiele parameters versus the normalized skyrmion radius ($R/\Delta$).  (a)  Gyrotropic $n$ and dissipation $d$ parameters.
    (b) Spin Hall effect force related length $f$ (divided by $\Delta$ so that is represented as a function of $R/\Delta$).
    (c) $n/d$, proportional to the skyrmion deflection $\rho$.
    (d) $f/d$, proportional to the skyrmion deflectionless velocity $v_0$.
    The solid lines correspond to the analytical expressions in the text and the dots correspond to the simulations of a GdCo ferrimagnetic alloy at different fields and temperatures.
    Note that $\Delta$ is not constant (Fig.~\ref{fig:radius}).
    In c) and d), the dashed line shows the small skyrmion limit ($n/d=1$, $|f/d|=\pi R/2$) and the dotted line the large skyrmion limit ($n/d=2\Delta/R$, $|f/d|=\pi\Delta/2$).
}
\label{fig:nfl}
\end{figure}

A strong variation of the dynamics with the geometric parameters in eqs.~\ref{eq:Thiele parameters ndf} is expected when the skyrmion radius is comparable to the domain wall width parameter
$\Delta = \sqrt{A/K_{\mathrm{eff}}}$ (with $A$ the exchange stiffness and $K_{\mathrm{eff}}$ the effective anisotropy). This is important near $T_R$, as in the experiments, where $\Delta$ can be large.
Since $n$ corresponds to the skyrmion topological number, it is independent of the skyrmion size ($n = \pm 4\pi$ depending on the skyrmion core polarity).
The integral $d$ is similar to the micromagnetic exchange integral~\cite{Thiele1974,vakili2021}, and it is expected to scale likewise with the skyrmion size.
The integral that defines $f$ is similar to a micromagnetic Dzyaloshinskii-Moriya integral~\cite{Hellman2017a} and therefore involves the skyrmion chirality $c$ ($c = \pm1$ respectively for clockwise and counterclockwise spin rotation).
In the limit of $R\gg\Delta$, $f \rightarrow \pi^2cR$. Since $f\rightarrow 0$ for small skyrmions, this linear variation can be used as an approximation for the full range of size. We also note that it is independent of $\Delta$. The variation of these three parameters with the skyrmion radius is shown by the lines in Fig.~\ref{fig:nfl}a,b.

For a skyrmion radius $R$ large as compared to $\Delta$, $d = 2\pi R/\Delta~$~\cite{Sampaio2013a,Hrabec2017c}. However, as shown by Belavin and Poliakov \cite{Belavin1975}, an exchange integral does not vanish at small size for a topological texture~\cite{Belavin1975}, and, for $R\rightarrow 0$, $d \rightarrow |n|$, which shows that the dissipation does not vanish at small sizes~\cite{Buttner2018}.
To describe the dissipation over the full range of $R$, the two limits can be interpolated as
\begin{equation}\label{eq:f_model}
    d \approx|n|\exp\left(-\frac{2\pi R}{|n|\Delta}\right)+\frac{2\pi R}{\Delta}.
\end{equation}
This approximation is shown as dashed lines in Fig.~\ref{fig:nfl}. As a consequence, the mobility, which is proportional to $f/d$ (Eq.~\ref{eq:ThieleSolEffPars}), vanishes for small skyrmions (Fig.~\ref{fig:nfl}d), and the deflection is maximum but does not diverge (Fig.~\ref{fig:nfl}c).

\section{Micromagnetic simulations \label{app:Simus}}

We performed micromagnetic simulations of skyrmions in a GdCo thin film using MuMax$^3$~\cite{vansteenkiste2014}, modified to account for the specificity of ferrimagnetic films~\cite{Haltz2020,Haltz2021,SI}, each lattice being described independently, and coupled with an antiferromagnetic coupling $J_\mathrm{TM/RE}$.
The parameters used for the simulations (Table~\ref{tab:parameters}) are those corresponding the skyrmions phase in Fig.~\ref{fig:charact}. We keep all parameters constant in temperature except for the sublattice magnetization and angular momentum ($M_S^{\rm Co}(T)$, $M_S^{\rm Gd}(T)$ calculated with the mean field model, Fig.~\ref{fig:charact}a; $L_S^{\rm Co}=M_S^{\rm Co}(T)/\gamma_{\rm Co}$, $L_S^{\rm Gd}=M_S^{\rm Gd}(T)/\gamma_{\rm Gd}$).

\begin{table}
    \caption{
        Parameters used for the micromagnetic simulations. Only the $M_S$ and $L_S$ of the two sublattices were considered to vary with temperature according to the mean field model (Fig.~\ref{fig:charact}a). Their range is shown in brackets. $J_\mathrm{TM/RE}$ is the interlattice coupling, and $g$ is the g-factor ($\gamma=g\mu_B/\hbar$, where $\mu_B$ is the Bohr magneton). }
    \label{tab:parameters}
    \begin{ruledtabular}
    \begin{tabular}{l l l}
                                        &   Co          & Gd\\  \hline
      $g$                     &   2.22        & 2.00\\
      $\alpha$          & 0.019 & 0.019 \\
      $M_S$ (MA/m)                      & [0.62 -- 0.5] & [1.1 -- 0.4] \\
      $K$  (kJ/m$^{3}$)                    & 11.5   & 0\\
      $A$ (pJ/m)    & 4.6    & 0\\
      $D_\mathrm{DMI}$ (mJ/m$^{2}$) & 0.22 & 0\\
      $J_\mathrm{TM/RE}$ (MJ/m$^{3}$)    & \multicolumn{2}{c}{25} \\
      $\theta_{\rm SHE}$       & 0.03     & 0\\
    \end{tabular}
    \end{ruledtabular}
\end{table}

\begin{figure}[ht]
    \includegraphics[width=\columnwidth]{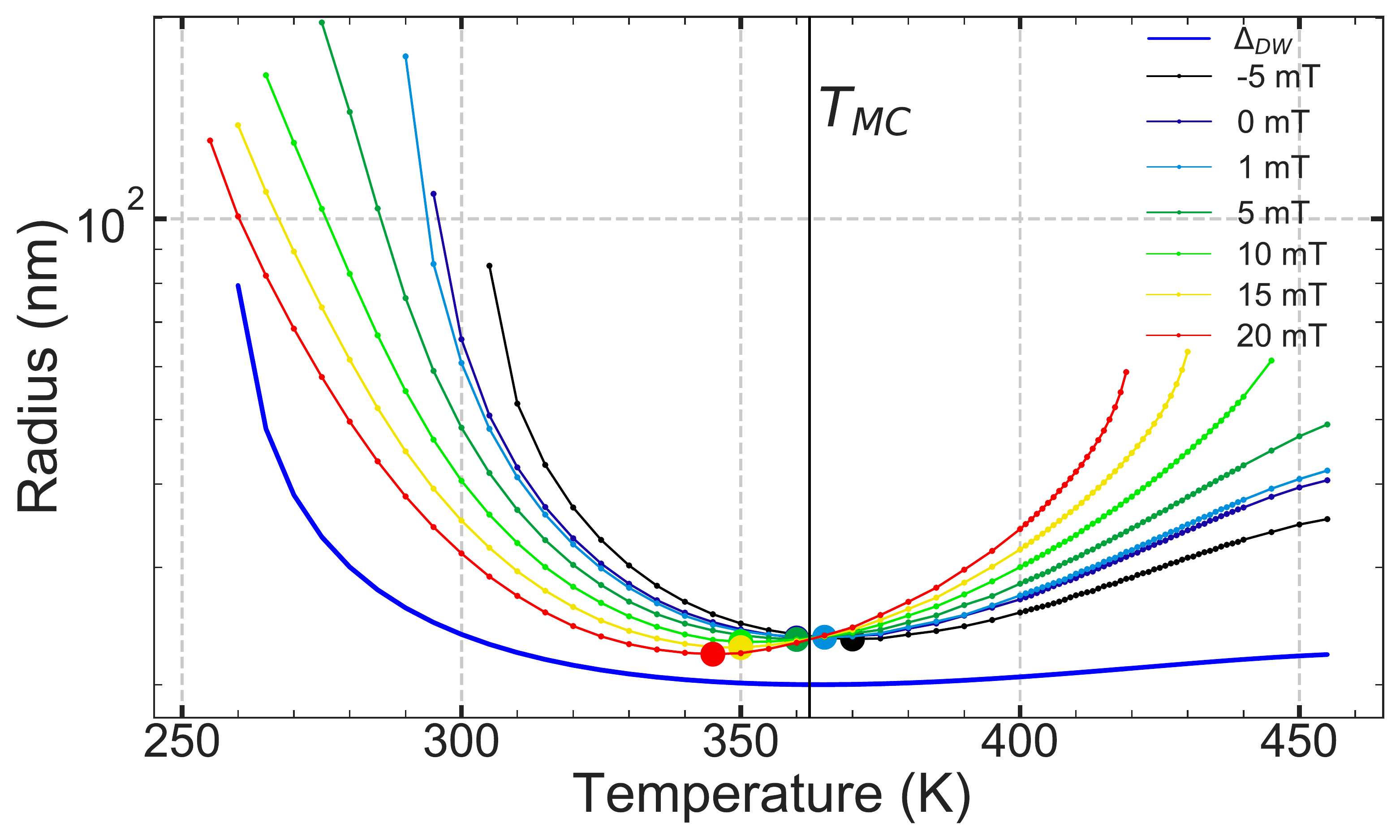}
    \caption{ Skyrmion radius versus temperature at several magnetic fields. The circles represent the minimum radius of each field.
    }
    \label{fig:radius}
\end{figure}

The simulated skyrmion radius versus temperature shows a minimum at \TMC{} (Fig.~\ref{fig:radius}). As observed experimentally, the external field changes the size of the skyrmion. This effect is opposite above and below \TMC{} as the net magnetization of the skyrmion core changes sign. The temperature of the minimum radius is, therefore, shifted by the field.

\begin{figure}[ht]
    \includegraphics[width=0.90 \columnwidth]{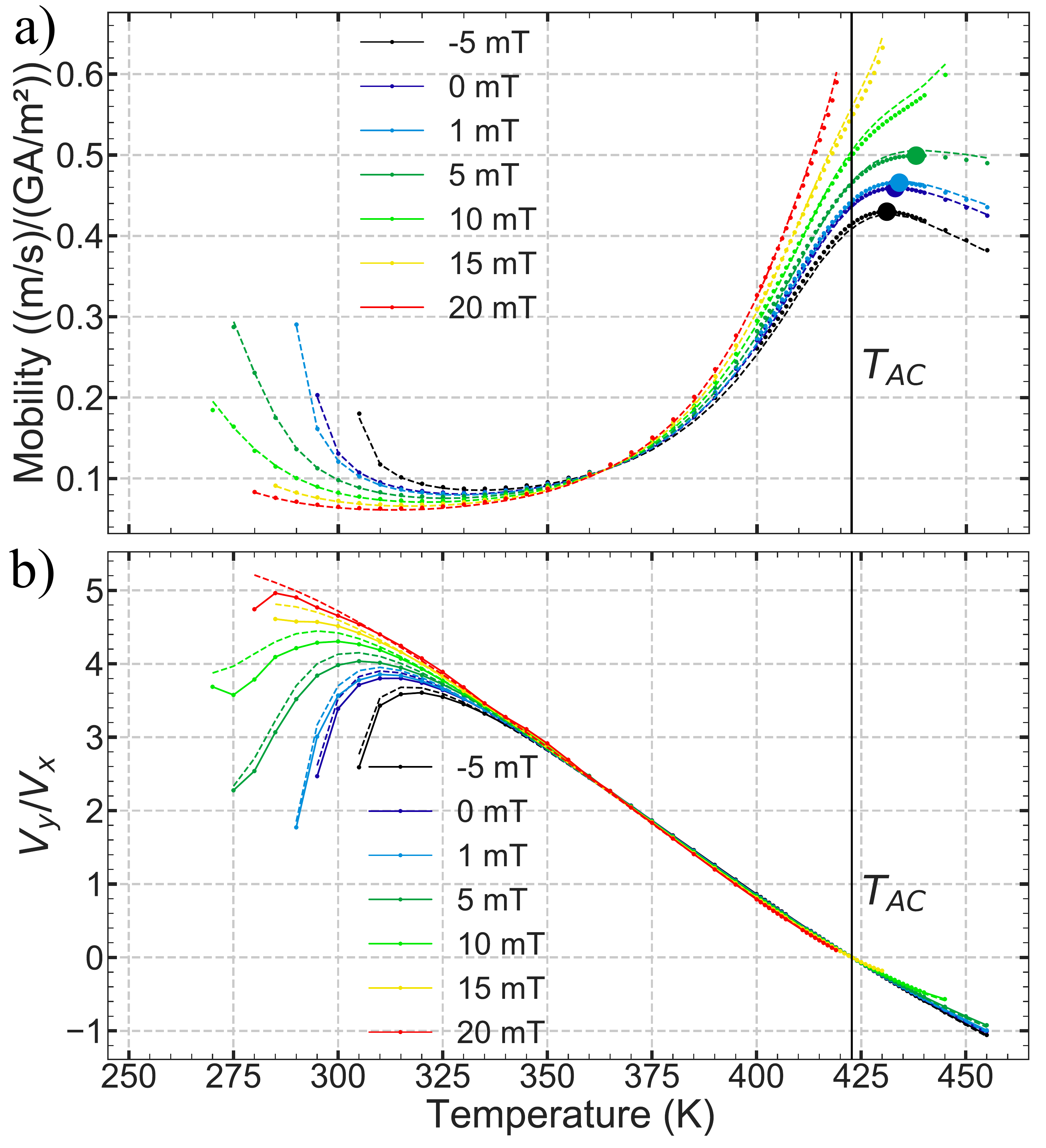}
    \caption{
        (a) Mobility $|v|/J$ versus temperature for different magnetic fields from simulations (points) and model (dashed lines). The circles represent the maximum mobility for each field.
        (b) deflection $v_y/v_x$ versus temperature and field. $J=100$~GA/m$^2$.
    }
    \label{fig:v_free}
\end{figure}

The skyrmion velocity versus temperature and field is shown in Fig.~\ref{fig:v_free}a. These simulations are performed in a low current regime  ($J=100$~GA/m$^2$) so that does not deform the skyrmion. A velocity maximum is observed close to \TAC{}, as predicted by the model.
However, the velocity curve is more complex than a simple peak. Since the mobility depends on the radius, and therefore on the temperature, the result is affected by both the skyrmion size and the angular momentum compensation. The mobility minimum at \TMC{} is due to the size minimum (Fig.~\ref{fig:radius}). The maximum observed slightly above \TAC{} is due to a combination of vanishing $L_S$ and the increasing size with temperature.
The model, using no fitting parameters other than the static skyrmion diameter (obtained from simulations), is in quantitative agreement with the simulations, a consequence of the simulated skyrmion conserving its static profile during motion and despite the finite coupling (the case of imperfect anti-parallel alignment is treated in \cite{SI}.)
The skyrmion deflection $v_y/v_x$ crosses zero at \TAC{} and is linear around \TAC{}. This supports the model which predicts that $v_y/v_x \propto L_S$.
Far from \TAC{}, the deflection deviates from this linear relation due to the large skyrmion size, as the deflection decreases as $1/R$.

\section{Material parameters \label{app:Params}}
We determined some of the material parameters using Brillouin light scattering measurements \cite{SI,Haltz2022}: exchange stiffness $A = 4.6$~pJ/m, the DMI parameter $D_{\rm DMI} = -0.22$~mJ/m$^2$ (which yields a surface DMI parameter $D_S = 1.1$~pJ/m), $\gamma/2\pi = 18.3$~GHz/T and $\alpha = 0.15$. The found exchange stiffness $A$ was found to be in agreement with \cite{Eyrich2014,Katayama1978}.
We can evaluate how these measurements compare to the expected effective values given by Wangsness theory.
The value of $\gamma$ is much lower than the value of Co, as expected from Wangsness formula.
We expect that the measured effective $\alpha$ be larger than the $\alpha$ of the Co sublattice, which is indeed the case \cite{Devolder2013}.

Table~\ref{tab:CharactPars} shows the calculated $l_C$ and critical DMI parameter $D_C = 4 \sqrt{A K_{eff}}/\pi$ at which stripes are favored,
calculated for three temperatures.

\begin{table}[ht]
    \centering
    \begin{ruledtabular}
 \begin{tabular}{l l l l}
                    & 280 K     & 290 K     & 300 K\\ \hline
    $M_S$ (kA/m)    &  90       & 78        &65  \\
    $K_\eff$ (kJ/m$^3$) & 6.4   & 7.7       & 8.8 \\
    $l_{c}$ (nm)    & -0.4      & 8         & 22 \\
    $D_{c}$ (mJ/m$^2$)& 0.22 & 0.24 & 0.26\\
    $\Delta$ (nm)   &  27       & 24    & 23    \\
 \end{tabular}
 \end{ruledtabular}
\caption{ Sample magnetic parameters.
$M_S$ measured by SQUID (Fig.~\ref{fig:charact}a), $H_{k}$ extracted from AHE measurements~\cite{SI}. The dipolar characteristic length is $l_{c}=\frac{\sigma}{\mu_{0} M_{s}^2}$, where $\sigma=4\sqrt{A K_\eff} - \pi D_{\rm DMI}$ is the domain wall energy.}
 \label{tab:CharactPars}
\end{table}


%

\end{document}